\documentclass[prl,reprint,amsmath,amssymb,aps]{revtex4-1}
\pdfoutput=1
\usepackage{geometry,enumerate,amsmath,amssymb}
\usepackage{verbatim}
\usepackage{fullpage}
\usepackage{graphicx}
\usepackage{dcolumn}
\usepackage{bm}
\newcommand{\be}{\begin{equation}}
\newcommand{\ee}{\end{equation}}
\newcommand{\bea}{\begin{eqnarray}}
\newcommand{\eea}{\end{eqnarray}}

\renewcommand{\epsilon}{\varepsilon}

\begin{document}
\title{Skyrmions and clustering in light nuclei}
\author{Carlos Naya and Paul Sutcliffe}
\affiliation{
Department of Mathematical Sciences,
Durham University, Durham DH1 3LE, United Kingdom.\\ 
Email:
carlos.naya-rodriguez@durham.ac.uk, \ p.m.sutcliffe@durham.ac.uk}
\date{October 2018}

\begin{abstract}
  One of the outstanding problems in modern nuclear physics is to determine the properties of nuclei from the fundamental theory of the strong force, quantum chromodynamics (QCD).  Skyrmions offer a novel approach to this problem by considering nuclei as solitons of a low energy effective field theory obtained from QCD. Unfortunately, the standard theory of Skyrmions has been plagued by two significant problems, in that it yields nuclear binding energies that are an order of magnitude larger than experimental nuclear data, and it predicts intrinsic shapes for nuclei that fail to match the clustering structure of light nuclei. Here we show that extending the standard theory of Skyrmions, by including the next lightest subatomic meson particles traditionally neglected, dramatically improves both these aspects. We find Skyrmion clustering that now agrees with the expected structure of light nuclei, with binding energies that are much closer to nuclear data.
\end{abstract}
\maketitle

QCD is the fundamental theory of the strong nuclear force and describes how quarks are confined to form protons and neutrons, together with the binding of these nucleons to form atomic nuclei. However, the complexity of the non-perturbative regime means that extracting the properties of nuclei directly from QCD is not within reach of current computational capabilities. Traditional methods of nuclear physics have confirmed that protons and neutrons are excellent effective degrees of freedom at the nuclear energy scale, but establishing a link to the more fundamental theory will not only provide a more complete understanding of nuclear physics but will also allow predictions for experimentally unknown nuclei and for matter under extreme conditions, for example in the interior of neutron stars.

Skyrmions are named after the British physicist Tony Skyrme, who introduced the standard version of the model almost sixty years ago \cite{Sk} as a nonlinear field theory of the lightest subatomic meson particles, called pions. This theory has topological soliton \cite{book} solutions, that is, twisted localized particle-like excitations of the pion fields, that are now known as Skyrmions. The number of twists in the pion fields corresponds to the number of Skyrmions and Skyrme proposed that this be identified with baryon number: which is equal to the mass number $A$, that counts the number of nucleons in a nucleus. This proposal was verified twenty years later \cite{Wi} by demonstrating that the model may be regarded as a low energy effective field theory of QCD in the limit of a large number of quark colours.
Skyrmions therefore provide an intermediate approach to nuclei, between the currently intractable fundamental theory of quarks, and the accuracy of more conventional nuclear physics methods based directly on protons and neutrons.
\begin{figure*}[ht]\begin{center}
 \includegraphics[width=15cm]{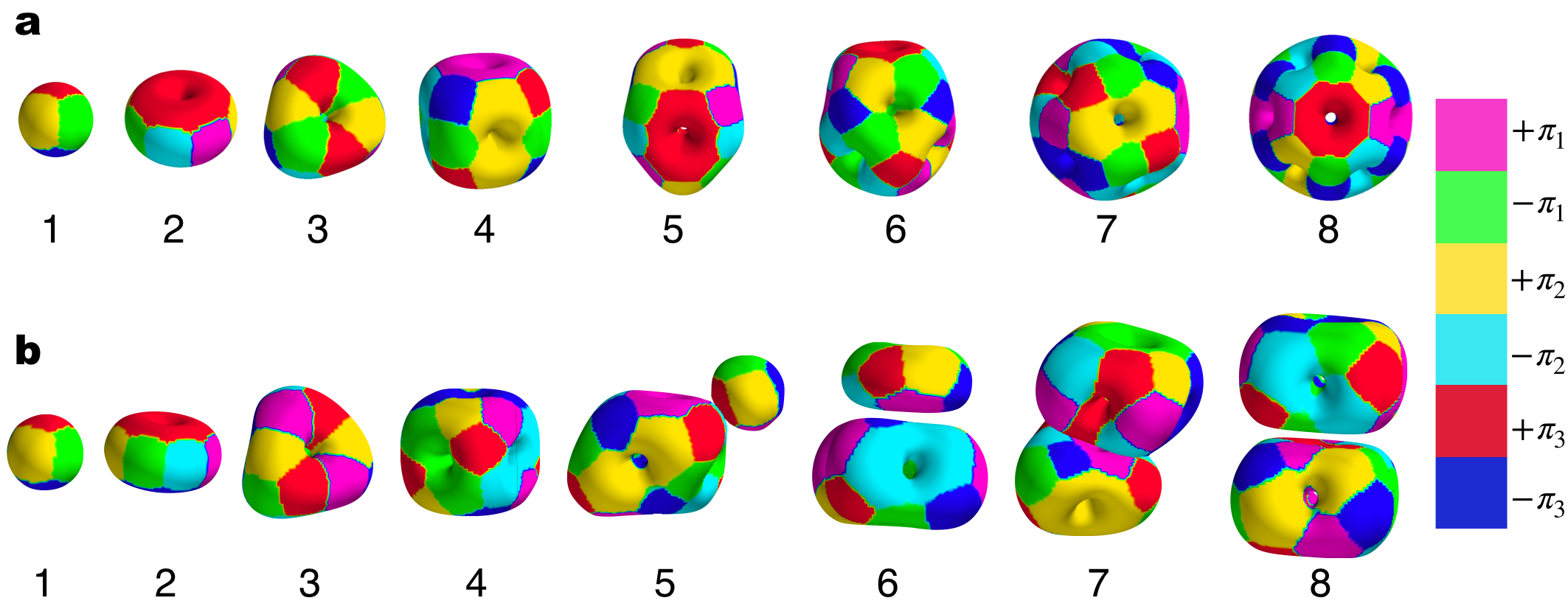}
 \caption{
   Baryon density isosurfaces for Skyrmions with nucleon numbers $A=1$ to $A=8$:
   (a) in the standard Skyrme model,  (b) in the extended Skyrme model that includes both pions and rho mesons. Colours indicate which of the three constituent pion fields has the largest magnitude and its sign, according to the colouring scheme shown.   
 }
\label{fig1}
\end{center}\end{figure*}

Skyrmions in the standard version of the model are displayed in Fig.1a, for nucleon numbers $A=1$ to $A=8$, by plotting baryon density isosurfaces that reveal the intrinsic shapes of the nuclei predicted by these Skyrmions \cite{BS}.
Although Skyrmions have had some success in modelling nuclei over the last few decades \cite{RZ} there are two major problems with Skyrme's original version of the theory. The first is that it produces binding energies for nuclei that are an order of magnitude larger than nuclear data obtained from experiments, and the second is that it does not reproduce the clustering structure of light nuclei suggested by both experimental data and more conventional nuclear theories \cite{Fr}.

The novel aspect of Skyrmions is that nuclei composed of baryons miraculously appear as solitons in a field theory of meson particles. In the standard version of the Skyrme model only the lightest meson particles, pions, are included within the theory. Heavier mesons are expected to provide corrections to this leading order theory but they are neglected in the standard Skyrme model, simply to make the computation of Skyrmions tractable.  By performing extensive parallel computations on a high performance computing cluster, we have been able to obtain the first results for Skyrmions in the theory containing both massive pions and rho mesons, the next lightest of the meson particles. In this letter we show that including the previously neglected rho mesons dramatically improves the two major failings of the standard Skyrme model highlighted above. Namely, Skyrmions now produce the required cluster structure of light nuclei, with binding energies that are much closer to nuclear data.

In the standard version of the Skyrme model \cite{Sk} the triplet of pion fields $(\pi_1,\pi_2,\pi_3)$ is encoded in the $SU(2)$-valued Skyrme field
\be
U=\begin{pmatrix}
\sigma +i\pi_3 & i\pi_1+\pi_2 \\ i\pi_1-\pi_2 & \sigma -i\pi_3
\end{pmatrix},
\ee
where the auxillary sigma field imposes the constraint $\sigma^2+\pi_1^2+\pi_2^2+\pi_3^2=1.$
The three
 $\mathfrak{su}(2)$-valued currents are defined to be $R_i=\partial_iU\,U^{-1}$ and in dimensionless units the static energy that defines the Skyrme model is given by
\begin{multline}
E_\pi=\int \bigg(
-\frac{c_1}{2}\mbox{Tr}(R_iR_i)-\frac{c_2}{16}\mbox{Tr}([R_i,R_j]^2)\\
+\frac{m^2c_1^2}{c_2}\mbox{Tr}(1-U))
\bigg)\,d^3x.
\label{enpion}
\end{multline}
Without loss of generality, the positive constants $c_1$ and $c_2$ can be set to unity by rescaling the dimensionless energy and length units, and this choice of scaling is known as using Skyrme units. However, in this study it will be convenient to work with a different scaling, so we set $c_1=0.141$ and $c_2=0.198$, to match with the normalization of the extended version of the model to be introduced later. The constant $m$ is the pion mass in dimensionless Skyrme units and is fixed by the experimental value.

The only parameters of the Skyrme model are therefore the two conversion factors to convert dimensionless energy and length units into physical units. Two physical quantities are required as input to determine these two factors and the common practice is to use the conversion values calculated by fitting to the properties of the proton and its excited state the delta baryon \cite{AN}. In these units the physical pion mass corresponds to the value $m=0.526$, which we take from now on. 

Baryon number is identified with the integer-valued topological charge
\be
B=\int \frac{1}{24\pi^2}\varepsilon_{ijk}\mbox{Tr}(R_iR_kR_j)\, d^3x,
\label{baryon}
\ee
with the integrand being the baryon density ${\cal B}$. Skyrmions that model nuclei with mass number $A$ are the stable energy minima of the energy $E_\pi$ with $B=A$.

Skyrmions are computed, as described in detail in previous work \cite{BS2}, by evolving dynamical second-order in time field equations derived from a Lagrangian with a static contribution equal to $-E_\pi,$ where fourth-order accurate finite difference approximations are used to evaluate spatial derivatives on a cubic lattice with boundary condition $U=1$. Flow to minimal energy states is achieved by instantaneously freezing the motion, via setting all time derivatives to zero, whenever $E_\pi$ is increasing. The simulations presented in Fig.1 were performed on a cubic lattice containing $128^3$ lattice points with a lattice spacing $\Delta x=0.08$ and the time evolution implemented via a fourth-order Runge-Kutta method with a timestep $\Delta t=0.02$.

The results for Skyrmions in the standard Skyrme model are presented in Fig.1a by plotting baryon density isosurfaces ${\cal B}= 0.02.$ These surfaces are coloured to indicate which of the three constituent pion fields has the largest magnitude and its sign, according to the colouring scheme shown in the figure. This information is relevant for understanding the forces between Skyrmions, because the pion fields interact as a triplet of orthogonal dipoles.

\begin{figure}[h!]\begin{center}
 \includegraphics[width=\columnwidth]{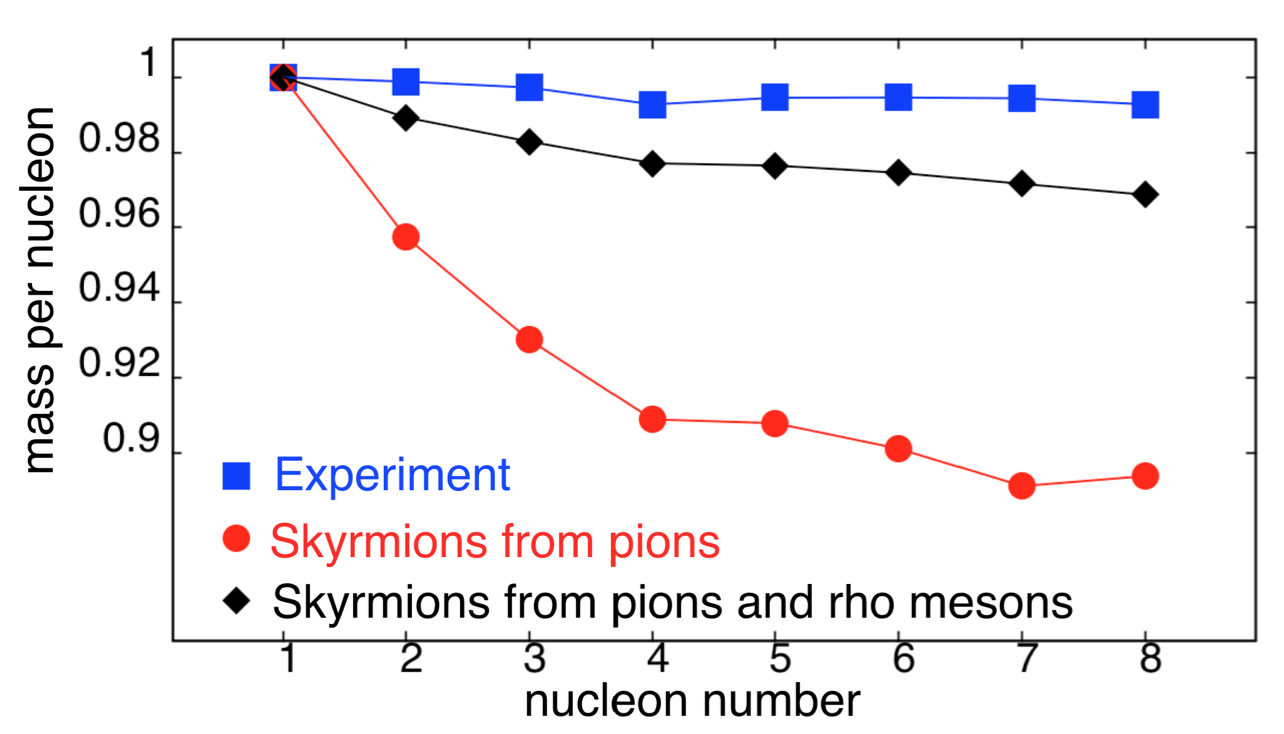}
 \caption{Experimental nuclear data on the mass per nucleon, in units of the proton mass (blue squares). The mass per nucleon of Skyrmions in the standard version of the Skyrme model of pions (red circles) and in the extended version of the Skyrme model including both pions and rho mesons (black diamonds),  in both cases normalized by the single Skyrmion mass.}
\label{fig2}
\end{center}\end{figure}
The data presented in Fig.2 highlights the problem with Skyrmion binding energies in the standard Skyrme model. The blue squares show the experimental nuclear data on the mass per nucleon, in units of the proton mass, for nuclei with nucleon numbers $A=1,..,8.$ This demonstrates that the binding energies of nuclei are no greater than 1\% of the mass of the nucleus. The red circles denote the mass per nucleon of Skyrmions $E_\pi/A$, normalized by the single Skyrmion mass.  In contrast to the experimental data, this plot confirms that the binding energy of a Skyrmion can be greater than 10\% of it’s mass: an order of magnitude larger than experimental values.

Despite this quantitative failing, the intrinsic shapes of several of the Skyrmions shown in Fig.1a are known to have some promising features. In particular, to interpret the classical Skyrmion solutions as nuclei requires the introduction of spin and isospin, which is generally performed via a semi-classical quantization that treats the Skyrmion as a rigid body that is free to rotate in both space and isospace. The symmetries of the classical Skyrmion solutions determine the allowed spin and isospin states and these have been calculated for many Skyrmions \cite{Kr}, including all those shown in Fig.1a. The predicted ground state spins and isospins for nucleon numbers $A=1,2,3,4$ match with experiment as a result of the symmetries of these Skyrmions. Unfortunately, the above match between Skyrmion states and nuclear data begins to break down at $A=5$ and is particularly poor for odd values of $A$. This signals a problem with the intrinsic shapes of Skyrmions for $A>4$ that is not unexpected and can already be anticipated from the images in Fig.1a. As discussed in more detail below, the basic problem with Skyrmions for $A>4$ is that they fail to show a cluster structure, and instead allow all constituents to merge and form configurations that are too symmetric.  

The extended Skyrme model requires the inclusion of the three $\mathfrak{su}(2)$-valued rho meson fields $\rho_i.$ These are introduced using the dimensional deconstruction formulation \cite{Su1}, that has the advantage of introducing no additional free parameters, and yields the energy \cite{Su2} of the extended model given by $E_{\pi,\rho}=E_\pi+E_\rho+E_{\rm int}$, where
\begin{multline}
E_\rho=\int -\mbox{Tr}\bigg\{
\frac{1}{8}(\partial_i \rho_j-\partial_j \rho_i)^2
+\frac{1}{8}\rho_i^2 \\
+c_3(\partial_i \rho_j-\partial_j \rho_i)[\rho_i,\rho_j]
+c_4[\rho_i,\rho_j]^2
\bigg\}\,d^3x,
\label{enrho}
\end{multline}
\begin{multline}
E_{\rm int}=\int -\mbox{Tr}\bigg\{
c_5([R_i,\rho_j]-[R_j,\rho_i])^2\\
-c_6[R_i,R_j](\partial_i \rho_j-\partial_j \rho_i)
-c_7[R_i,R_j][\rho_i,\rho_j]
\\
+\frac{1}{2}c_6[R_i,R_j]([R_i,\rho_j]-[R_j,\rho_i])\\
-\frac{1}{8}([R_i,\rho_j]-[R_j,\rho_i])(\partial_i \rho_j-\partial_j \rho_i)
\\
-\frac{1}{2}c_{3}([R_i,\rho_j]-[R_j,\rho_i])[\rho_i,\rho_j]
\bigg\}\,d^3x,
\label{enint}
\end{multline}
with the values of the constants 
$c_3=0.153,\
c_4=0.050,\
c_5=0.038,\
c_6=0.078,\
c_7=0.049.$

Skyrmion solutions of the extended model are displayed in Fig.1b for nucleon numbers $A=1,..,8$, by plotting baryon density isosurfaces ${\cal B}=0.02.$
These Skyrmions were computed by applying the simulation scheme described above to minimize the extended energy $E_{\pi,\rho}$, with the condition $\rho_i=0$ imposed at the boundary of the simulation lattice.
These plots reveal that the intrinsic shapes of Skyrmions have completely changed for $A>4$ and now display the cluster structure expected of light nuclei, as discussed below. Furthermore, the black diamonds in Fig.2 show the mass per nucleon $E_{\pi,\rho}/A$ of Skyrmions in the extended model, again normalized by the single Skyrmion mass. This data reveals that binding energies are reduced from over 10\% to no more than 3\%, significantly improving the comparison with experimental data.

Both experimental evidence and a variety of theoretical  approaches support the existence of clustering in light nuclei \cite{Fr}, namely the emergence of molecular-like sub-units.  Clustering begins at $A=5$, where the nucleus of $^5\mbox{He}$ is regarded as an $\alpha$-particle ($^4\mbox{He}$) core with an orbiting neutron \cite{PTW1}. This is not reflected in the $A=5$ Skyrmion in the standard model, where all five Skyrmions are democratically merged to form a single structure. However, clustering is clearly evident in the $A=5$ Skyrmion in the extended model including rho mesons, where a single $A=1$ Skyrmion is isolated from a core that has the shape of a slightly deformed cube corresponding to the $\alpha$-particle of the $A=4$ Skyrmion.

Similar comments apply to the $A=6$ and $A=7$ Skyrmions, where neither the $\alpha$-particle plus deuteron ($^2\mbox{H}$) cluster \cite{PTW2} of $^6\mbox{Li}$ nor the $\alpha$-particle plus triton ($^3\mbox{H}$) cluster \cite{PTW2} of $^7\mbox{Li}$ are reflected in the intrinsic shapes of the standard Skyrmions, but are clear in the extended model with rho mesons. In fact the $A=7$ Skyrmion in the standard model is embarrassingly symmetric, having dodecahedral symmetry that predicts a ground state with a spin far greater than that seen in the experimental data for the ground state of $^7\mbox{Li}$ . Recently, an approach to address this failure has been proposed \cite{Ha} by including vibrations that encourage the too symmetric $A=7$ Skyrmion to split into the required $A=4$ plus $A=3$ cluster structure, but here we find that including rho mesons already yields this clustering without the need for vibrations.

Most attention on clustering has been directed towards the study of $\alpha$-particle sub-units in $\alpha$-conjugate nuclei, composed of an equal and even number of protons and neutrons \cite{Fr}. The first example is the 2$\alpha$ system of $^8\mbox{Be}$, but again the $A=8$ Skyrmion in Fig.1a shows no sign of reflecting this property. However, it has been found that using an artificially large pion mass, of around twice the physical value, does change the intrinsic shape of the $A=8$ Skyrmion to generate a 2$\alpha$ cluster, and also yields $N\alpha$ clusters for $A=4N$ Skyrmions \cite{BMS}. Despite this encouraging development in the standard version of the model with a large pion mass, this does not change the shapes of any of the Skyrmions with $A<8$ or reduce the large binding energies. Here we find that the 2$\alpha$ cluster of the $A=8$ Skyrmion with rho mesons, displayed in Fig.1b, appears without the need to artificially increase the pion mass, and indeed the cluster structure of a pair of $\alpha$-particles is now more obvious.

\begin{figure}[h!]\begin{center}
 \includegraphics[width=\columnwidth]{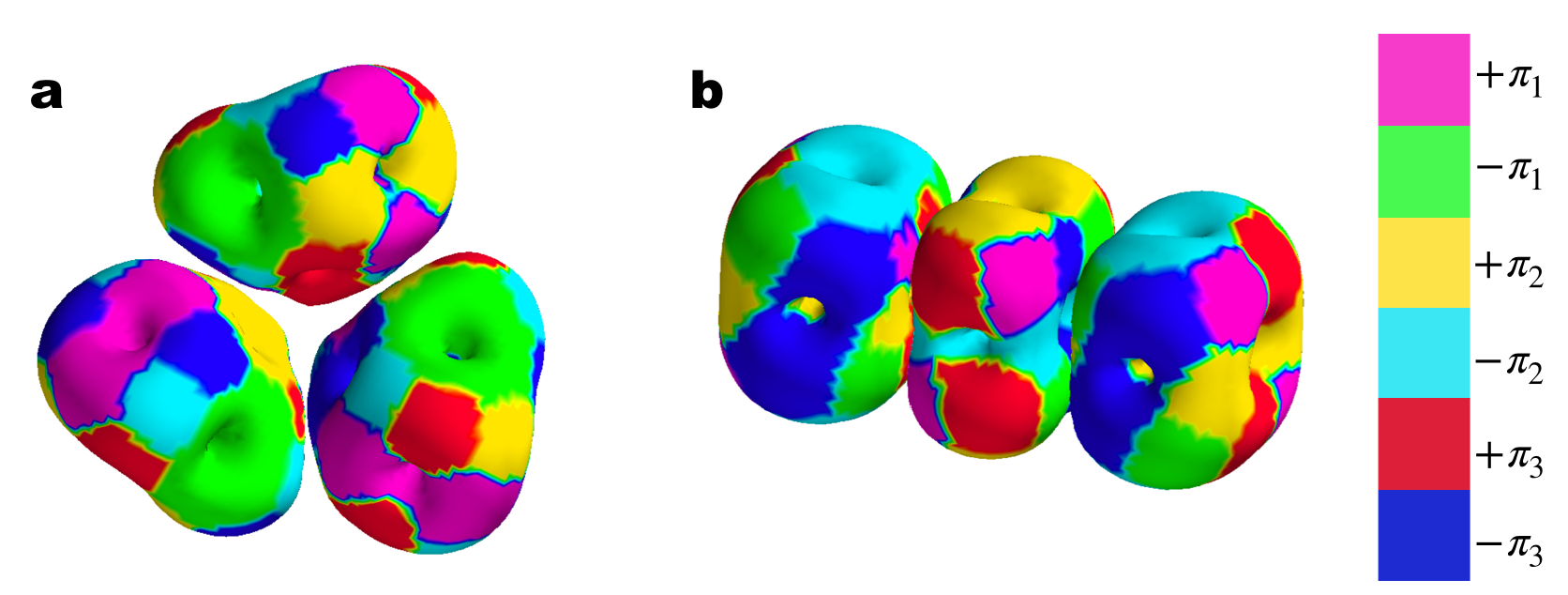}
 \caption{Baryon density isosurfaces for Skyrmions with $A=12$ in the extended Skyrme model:
   (a) $3\alpha$ triangular clustering, 
   (b) $3\alpha$ linear chain clustering.
   Colours indicate which of the three constituent pion fields has the largest magnitude and its sign, according to the colouring scheme shown.}
\label{fig3}
\end{center}\end{figure}

A success \cite{LM} of the standard Skyrme model, albeit it with a large value of the pion mass, is the next $\alpha$-conjugate nucleus of $^{12}\mbox{C}$, where there are two different $A=12$ Skyrmions, one with a triangular 3$\alpha$ structure and the other with a linear 3$\alpha$ arrangement, that have properties suggesting identifications with the ground state and the famous Hoyle state of $^{12}\mbox{C}$, respectively. Similar $A=12$ Skyrmions exist for both configurations in the extended model, see Fig.3, so this success of Skyrmions is maintained by the inclusion of rho mesons. In agreement with the situation in the standard Skyrme model with a large pion mass, we find that the linear chain cluster of Fig.3a has a slightly lower energy than the triangular cluster of Fig.3b.

In computing the Skyrmions displayed in Fig.1b a wide variety of initial conditions were applied for each value of $A>1$, to avoid trapping in local minima.
This included a product ansatz to generate $A$ initially separate single Skyrmions and the application of the rational map approximation \cite{HMS}, that provides a good description of Skyrmions in the standard Skyrme model. In each case there was a clear gap between the Skyrmions presented in Fig.1b and any other local energy minima, except for the case $A=6$, where another Skyrmion solution with an energy equal to the one shown (to within our expected numerical accuracy) was also obtained. This alternative solution also has a cluster structure, but rather than the $A=4$ Skyrmion plus the $A=2$ Skyrmion cluster shown in Fig.1b it is a cluster of two $A=3$ Skyrmions, arranged face-to-face to preserve a triangular symmetry.

In recent  work \cite{NS} Skyrmions of the extended model were computed in the simplifying limit in which the pions are assumed to be massless, that is, by minimizing the energy $E_{\pi,\rho}$ with $m=0.$ It must be stressed that this apparently innocuous simplification has dramatic consequences. In particular, there is no clustering behaviour in this limit and the Skyrmions retain the shapes of those in the standard Skyrme model.

In summary, we have shown that extending the standard version of Skyrmions, by including not only massive pions but also massive rho mesons, significantly improves the features of Skyrmions in exactly the areas where discrepancies with experimental results were most problematic, whilst retaining the successful aspects of Skyrmions. Of course, we do not expect the addition of rho mesons alone to now provide a perfect match between Skyrmions and nuclear data, because we have clearly shown that neglecting heavier mesons can have significant consequences. However, the results presented here provide considerable evidence that as each of the heavier mesons are included within the theory there is an optimism for the convergence of Skyrmions to nuclei.

\section*{Acknowledgements}
This work is funded by the
Leverhulme Trust Research Programme Grant RP2013-K-009, SPOCK: Scientific Properties Of Complex Knots, and by the European Union Horizon 2020 research and innovation programme under the Marie Sk\l odowska-Curie grant agreement No 702329. The parallel computations were performed on Hamilton, the Durham University HPC cluster.

\end{document}